\documentclass[letterpaper, 10 pt, conference]{IEEEtran}

\usepackage{url}

\usepackage{geometry}
\usepackage{graphicx}
\usepackage[latin1]{inputenc}
\usepackage{amssymb,amsmath,array}
\usepackage{multirow}
\graphicspath{{figures/}} 

\usepackage{color, soul}
\usepackage[final]{changes}

\usepackage{fancyhdr, lipsum}

\usepackage{dblfloatfix}

\definechangesauthor[name={1.1}, color=red]{1.1}

\fancypagestyle{titlepage}{
} %

\begin{document}
%

%
\title{\vspace{6mm} The OMG-Emotion Behavior Dataset}








\author{
\IEEEauthorblockN{Pablo Barros, Nikhil Churamani, Egor Lakomkin, Henrique Siqueira, \\Alexander Sutherland and Stefan Wermter} 
\IEEEauthorblockA{
Knowledge Technology, Department of Informatics\\
University of Hamburg, Hamburg, Germany\\
Email:\{barros, 5churama, lakomkin, siqueira, sutherland, wermter\}@informatik.uni-hamburg.de}
}


%

\maketitle
\thispagestyle{empty}
\pagestyle{empty}

\begin{abstract}

This paper is the basis paper for the accepted IJCNN challenge One-Minute Gradual-Emotion Recognition  (OMG-Emotion) \footnote{https://www2.informatik.uni-hamburg.de/wtm/OMG-EmotionChallenge/} by which we hope to foster long-emotion classification
using neural models for the benefit of the IJCNN community. The proposed corpus has as novelty the data collection and annotation strategy based on emotion expressions which evolve over time into a specific context. Different from other corpora, we propose a novel multimodal corpus for emotion expression recognition, which uses gradual annotations with a focus on contextual emotion expressions. Our dataset was collected from Youtube videos using a specific search strategy based on restricted keywords and filtering which guaranteed that the data follow a gradual emotion expression transition, i.e. emotion expressions evolve over time in a natural and continuous fashion. We also provide an experimental protocol and a series of unimodal baseline experiments which can be used to evaluate deep and recurrent neural models in a fair and standard manner.



\end{abstract}

\section{Introduction}
\label{sec:intro}

Affective systems have been used to categorize emotion expressions for the past two decades. Most of these systems are based on Paul Ekman's categorization scheme, known as the six universal emotions: ``Disgust'', ``Fear'', ``Happiness'', ``Surprise'', ``Sadness'', and ``Anger'' \cite{Ekman1971}. Although Ekman shows in his studies that these emotional categories are most commonly inferred from facial expressions, the way we express ourselves is more fluid and is, therefore, more difficult to categorize \cite{Hamann2012}. Humans usually express themselves differently, sometimes even combining one or more characteristics of the so-called universal emotions \cite{Izard1992}. This is somehow embedded in the dimensional emotion representation, usually described as arousal and valence spaces \cite{Thompson2011}.
    
Dealing with a set of restricted emotions, or a single instantaneous emotion description, is a serious constraint for most computational applications focused on any kind of human interaction \cite{Rani2006}. Humans have the capability of adapting their internal emotion representation to a newly perceived emotional expression on the fly and use the representation to obtain a greater understanding of another person's emotional behavior. This mechanism is described as a developmental learning process and after participating in different interactions, humans can learn how to describe traits of complex emotional behaviors such as sarcasm, trust, and empathy \cite{Thompson2011}.

Recent research trends in artificial intelligence and cognitive systems have approached computational models as a human-like perception categorization task. However, most of the research in the area is still based on instantaneous expression categorization where the task is to classify a single emotion expression using features from different modalities \cite{Soleymani2017}. This diverges from the developmental aspect of emotional behavior perception and learning, as instantaneous categorization lacks the contextual aspect of the data that comes from the continuous analysis of emotion expressions \cite{Ochsner2008}.

In recent years many corpora on what is known as emotion recognition ``in-the-wild" were released. All of these datasets, although very challenging, are focused on instantaneous emotion categorization. This means that a specific label is set for a short-term (usually a couple of seconds) emotion expression. There are corpora which have annotated interactions  \cite{Iemocap2008, mosi2016, emoreact2016}, they are however limited to restricted and limited-context scenarios, which do not allow the development of naturalistic emotion description models.

Researchers have previously performed studies on emotional behavior categorization and learning but most of them faced the problem of lacking a challenging and real-world-based corpus with long-term emotional relations that were annotated using a rigorous methodology. Thus, this paper proposes a dataset with richly annotated gradual emotional behavior categorization.

Our One-Minute-Gradual Emotion Dataset (OMG-Emotion Dataset) is composed of 567 relatively long emotion videos with an average length of 1 minute, collected from a variety of Youtube channels. The videos were selected automatically based on specific search terms related to the term ``monologue''. Using monologue videos allowed for different emotional behaviors to be presented in one context and that changes gradually over time. Videos were separated into clips based on utterances, and each utterance was annotated by at least five independent subjects using the Amazon Mechanical Turk tool. To maintain the contextual information for each video, each annotator watched the clips of a video in sequence and had to annotate each video using an arousal/valence scale and a categorical emotion based on the universal emotions from Ekman.

Each annotator was also given the full contextual information of the video up to that point when annotating the dataset. That means that each annotator could take into consideration not only the vision and audio information but also the context of each video, i.e. what was spoken in the current and previous utterances through the context clips provided by the annotation tool. In this manner, each annotation is based on multimodal information, different from most recent datasets on emotion recognition. This gives our corpus an advantage when used in cross-modal research, especially when analyzing the audio, vision, and language modalities.

We also present baseline emotion recognition results for the dataset. As our annotations are based on multimodal cues, we designed three experiments, one for each individual modality (vision, audio, and text). The baseline uses state-of-the-art neural models, based on deep learning architectures. Our initial results show that the corpus is very challenging and it is difficult for models to recognize emotions using an individual modality, and that there is space for improvement in future work that uses multimodal architectures.

By making our dataset publicly available, we intend to contribute to the neural network community to foster the development of neural models which can take into consideration multimodal contextual information, using techniques based on deep, recurrent and self-organizing networks.

\section{Related Work}

\begin{table*}
\caption{Summary of recent emotion expression corpora which are similar to the proposed OMG-Emotion dataset }
    \center \begin{tabular}{ |c | c |c| c| c| c| c| c|}
    \hline
    Dataset  & Modalities & Annotation Domain & Samples & Annotation Level & Annotators  & Annotation Strategy & Scenario\\ \hline
    IEMOCAP \cite{Iemocap2008} & A, V, L & Dimensional, 9 Emotions & 6000 & Utterance & 5 & Contextual & Indoor\\
    MOSI \cite{mosi2016} & A, V, L & Head gestures, Sentiment & 2199 & Video Clip& 5 per video & Instance & Wild\\
    EmoReact \cite{emoreact2016}& A, V & 17 Emotions & 1102 & Video Clip &1 per video& Instance & Wild\\
    GIFGIF+ \cite{Chenetal2017} & V & 17 Emotions & 25.544& Video Clip&500+ per video &Instance & Wild\\
    Aff-Challenge \cite{Zafeiriou2017} & A, V & Dimensional & 400 & videos & 1 & Contextual & Wild\\
    EMOTIW \cite{Dhall2017} & A, V & 7 Emotions & 1809  & Videos & 3 & Instance & Wild\\
    OMG-Emotion & A, V, L & Dimensional, 7 Emotions & 2400 & Utterance & 5 per video & Contextual & Wild\\
    \hline
    
    \end{tabular} 
\label{tab:datasetsComparison}
\end{table*}

In recent years, there has been an increased research interest in emotion expression corpora. This increase was mostly caused by advances in deep neural networks, and their applicability to real-world information. The new range of possible application areas also explains the increased popularity for datasets and solutions based on ``in-the-wild'' information, i.e. data that is extracted from scenarios which are very similar to, or directly from, the real world. Table \ref{tab:datasetsComparison} exhibits some of the most recent and popular corpora on emotion expression recognition.

The Interactive Emotional Dyadic Motion Capture (IEMOCAP) \cite{Iemocap2008} dataset is one of the most used corpora for training emotion recognition systems. It contains 12 hours of audiovisual recordings of dialogue sessions between two actors. Each dialogue session is split into utterances and each utterance is annotated by five different annotators. Annotators evaluated the arousal/valence and categorical emotion (Neutral, Happiness, Sadness, Anger, Surprise, Fear, Disgust, Frustration and Excitement) of each utterance. The annotators were instructed to watch and annotate the entire dialogue in sequence, in order to keep the contextual information in their annotations. This avoids improbable transitions in one dialogue (from frustrated to satisfied in a single sentence). Although this dataset is used by many researchers, it is heavily limited by the data collection method: The recordings happened in an indoor scenario, and although the subjects performed scripted and non-scripted dialogues, the instructions given to the subjects limited the types of emotions that can be expressed.

In the related field of sentiment analysis, the Multimodal Corpus of Sentiment Intensity and Subjectivity Analysis in Online Opinion Videos (MOSI) \cite{mosi2016} has in-the-wild data. This corpus contains 2199 video segments obtained from Youtube by a data crawler for the purpose of sentiment analysis. The authors of this corpus provide extensive annotations based on sentiment analysis of visual gestures and verbal information. The dataset was annotated using the Amazon Mechanical Turk tool, and for each video segment, five different annotators evaluate the video based on a scale from strongly positive to strongly negative, and four head gestures (smile, head nod, frown and head shake). Although a very challenging dataset, it does not contain contextual information, unlike the IEMOCAP, and thus cannot be used to train emotional models that make use of changes in contextual information. The use of real-world information makes it more suitable for generalized emotion recognition, but the lack of emotional annotation, be it categorical or dimensional, reduces the applicability of the corpus for training automatic emotion recognition systems. 

Similar to the MOSI dataset, the EmoReact \cite{emoreact2016} also contains videos from Youtube. The special feature of this dataset is that it contains videos from children of the Kids React Youtube channel. The dataset is annotated with 17 different affective states: the six basic emotions (Happiness, Sadness, Surprise, Fear, Disgust and Anger), neutral, valence intensity, curiosity, uncertainty, excitement, attentiveness, exploration, confusion, anxiety, embarrassment and frustration. This dataset was also annotated using the Amazon Mechanical Turk tool, with the annotators using a 1-4 Likert scale for each emotion, where 1 shows the absence of emotion and 4 shows the intense presence of the emotion. This dataset presents a general annotation scheme, and although it does not have a dimensional annotation, it has the intensity of different emotion categories. The videos contain children only, and thus this corpus is highly biased to how children express emotions. Also, the lack of contextual information makes it more suitable to be used on instance-based emotion classification.

Using a similar approach to collect data as the MOSI and EmoReact corpora, the GIFGIF+ dataset \cite{Chenetal2017} has video clip segments of emotional expressions. The difference between this dataset and others is that it provides only vision information of the expressions. Annotation was performed using an online self-developed platform in a crowd-sourced manner and currently, the dataset boasts more than 3 million annotations. The data is available in GIF format and not as videos which, according to the authors, provides a challenge on different frame rates per sequence. This dataset presents an impressive number of annotations per video clip, which contributes to the universal opinion on the annotation categories. However, the dataset presents unrealistic frame rates within their video clips, with the highest frame rate being 40 times more than the slowest one. Also evident is that, for only providing one modality, this dataset is restricted to vision-only applications.

The development of challenges based on automatic emotion recognition is also on the rise. Two of the most difficult of these challenges are the Emotions in the Wild (EMOTIW) \cite{Dhall2017} and Aff-Challenge \cite{Zafeiriou2017}. Both datasets are composed of videos which are considered ``in-the-wild'': the EMOTIW has clips from movies and the AFF-Challenge has different Youtube videos. There are some differences in how they were annotated: while the AFF-Challenge was annotated based on a dimensional arousal/valence scale, the EMOTIW was annotated based on the six universal emotions and neutral. The annotations on the AFF-challenge were made with a mechanical joystick which the annotator had to move while watching the video clip, determining the arousal and valence individually for each video clip. Although both datasets have many ``in-the-wild'' samples, most of them were not composed of contextual information. The AFF-challenge dataset annotates the videos in a way that the contextual information is taken into consideration, but the low number of annotators and the nature of the videos itself makes it difficult for an emotion recognition model to learn continuous information. This happens because the videos present in this dataset are mostly from people reacting to different things shown in the video such as other video clips, movies, and announcements. This means that there is little contextual information available and annotators must rely on vision as little emotion through and language in a consistent fashion.

Our dataset was designed based on the strengths of each of these corpora: it contains in-the-wild videos, a large number of different annotators and, most importantly, it contains videos where emotion expressions emerge and develop over time based on the monologued scenarios. 

\section{Data Acquisition and Pre-processing}
\label{sec:arch}
\par Youtube was used as a source for dataset samples and a special crawler was developed to filter potential video candidates. Since the goal was to have candidates with emotional content, several Youtube channels were manually preselected with collections of recorded monologues and acting auditions. Our hypothesis was that dataset samples selected from Youtube should contain ample variety in recording conditions, speakers and expressed emotions. Acted monologues are likely to have utterances displaying both neutral and salient emotions, depending on the setting of the monologue itself.
\par
\begin{figure}[t]
    \centering
    \includegraphics[height=10cm]{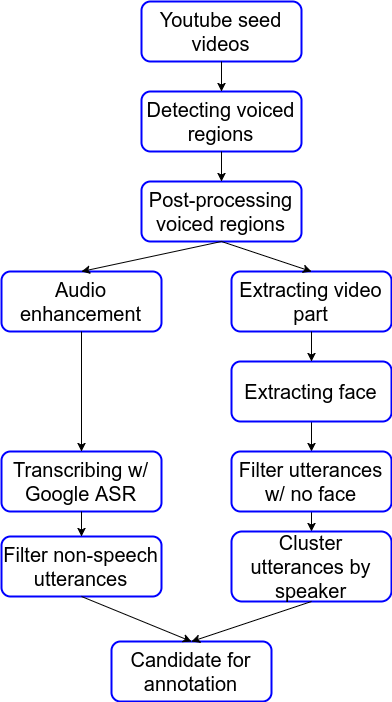}
    \caption{Flow chart illustrating the process of filtering and selecting of video samples.}
    \label{fig:crawler}
\end{figure}

 Several filtering steps were performed to select candidates (see Figure \ref{fig:crawler}). Firstly, the OpenSMILE toolkit \cite{eyben2013recent} was used to find regions with a high speech probability and a 0.5 probability value was used as the activation threshold. Voiced regions with a gap of less than 600 milliseconds between them were merged together, but maximum length of the utterance was capped at 10 seconds to prevent very long samples with many short pauses between sentences. Utterances were merged together if both were less then 500 milliseconds long. After the merging procedure utterances were discarded if they were shorter than one second. 

Video chunks were extracted according to the voiced regions using the ffmpeg\footnote{https://ffmpeg.org/ [Accessed 31.01.2018]} tool. Dlib\footnote{http://dlib.net/ [Accessed 31.01.2018]} was used to extract faces from each video frame. The volume of an utterance was normalized to 0 dB using SoX\footnote{http://sox.sourceforge.net/ [Accessed 31.01.2018]} and transcribed using the Google Speech Recognition system. Utterances were filtered out where multiple speakers were present in the video or if the transcribed spoken text was empty. Furthermore, we clustered together utterances in which the same person appeared sequentially, as this person was likely in the same context. Dlib was used to extract 128-dimensional facial embeddings and if the L-2 distance between two embeddings was greater than 0.6, faces were treated as if they came from different speakers. Finally, if the duration of the sequence exceeded 15 seconds and contained at least 3 utterances it was considered as a candidate for annotation.
\section{Annotation Strategy}

To annotate the collected data we built our own annotation tool, the KT Annotation Tool. This tool was designed to be a modular solution for media annotation, developed as a web platform. It was used with the Amazon Mechanical Turk solution and allowed us to collect a large number of unique annotations per video. In the following sessions, we describe the KT Annotation tool and our annotation procedure.

\subsection{KT Annotation Tool}

The KT-Annotation Tool is designed as a dynamic tool for collecting dataset annotations. It provides researchers with a web-based interface that can be tailored to the requirements of each dataset that needs to be annotated. A customizable front-end developed using the Django\footnote{https://www.djangoproject.com [Accessed 10.01.2018]} framework with a secure back-end built using SQLite\footnote{https://sqlite.org/ [Accessed 10.01.2018]} allows users to annotate images, text, audio samples or audio-visual content. The tool can be easily modified for different datasets using the administrative interface allowing researchers to easily collect corresponding annotations in a crowd-sourced manner. The ability to modify the front-end as well as the back-end functionality based on requirements posed by the dataset to be annotated provides an improvement over using pre-made non-modifiable templates provided by various existing annotation services. Furthermore, it provides a user-friendly alternative to writing a customized tool from scratch every time a new dataset needs to be annotated. 

\begin{figure}[t]
    \centering
    \includegraphics[width=\columnwidth]{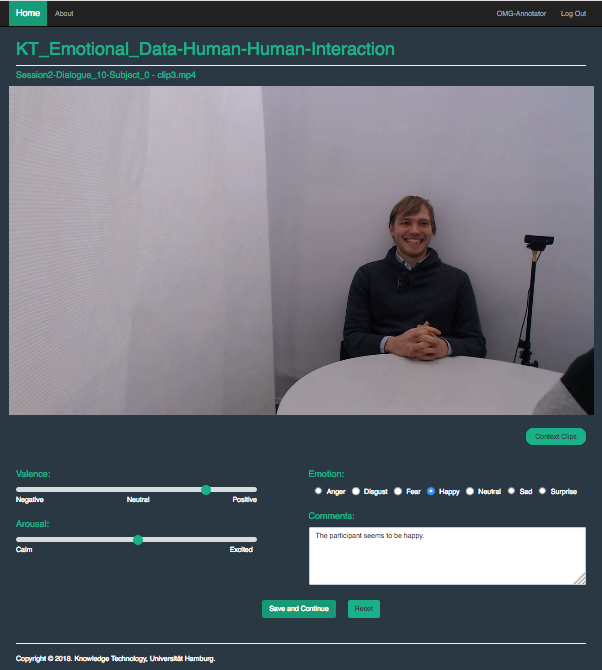}
    \caption{The User Interface of the tool used for annotations.}
    \label{fig:kttool}
\end{figure}

The tool provides a login service which can be used to provide customized access to multiple annotators at the same time. Once logged-in, annotators provide their evaluations on the data samples based on the parameters specified by the researchers. The labelling configuration used for annotating the OMG dataset can be seen in Fig.~\ref{fig:kttool} where a video sample is annotated for the affective context it represents. Each annotation is stored in the SQLite database allowing for efficient post-processing of the data. The recorded annotations can be exported in different formats such as comma-separated variables (.csv) or as spreadsheets (.xls) for data analysis and data distribution.

\subsection{Annotation Procedure}

For annotating the OMG dataset, a number of annotation tasks were published corresponding to each video in the dataset and the annotators were granted access to the annotation tool based on their MTurk ID and the unique IDs of the videos assigned to them. Each video was split into several clips based on utterances. The annotators provided their evaluations by setting valence, arousal and emotion labels for each clip individually. Valence and arousal values were recorded using two separate sliders, ranging from negative (-1) to positive extremes (1) for valence, as well as calm (0) and excited (1) for arousal. The intervals, [0,1] for arousal and [-1,1] for valence, are common representations found in most recent datasets. The annotators were not informed about these intervals, and they were calculated after the annotations were performed exclusively for computational purposes. Emotions were recorded using radio buttons representing seven categories corresponding to the six universal emotions~\cite{Ekman1971} and `Neutral'. Annotators were allowed to see the previous clips they had already annotated for a particular video by using the `Context Clips' option (Fig.~\ref{fig:kttool}) in order to maintain a consistent notion of the current data context. They were restricted from seeing and updating their annotations for these samples to avoid any bias. This was done such that each evaluation could be used individually as well as in the context of the entire video. For each video, we obtained 5 annotations from different annotators.

Once all the clips in the video sequence were successfully annotated, a unique success code was generated for the video. This success code was entered by the annotators to claim their remuneration in Amazon Mechanical Turk. If the annotator was logged out or unable to complete an annotation sequence, they could log back into the tool using the same video ID and the annotation process resumed from the same point in the sequence, ensuring a higher number of completed annotations. 

%
%



\section{Data Analysis}

\begin{figure*}[t]
    \centering
    \includegraphics[width=1\textwidth]{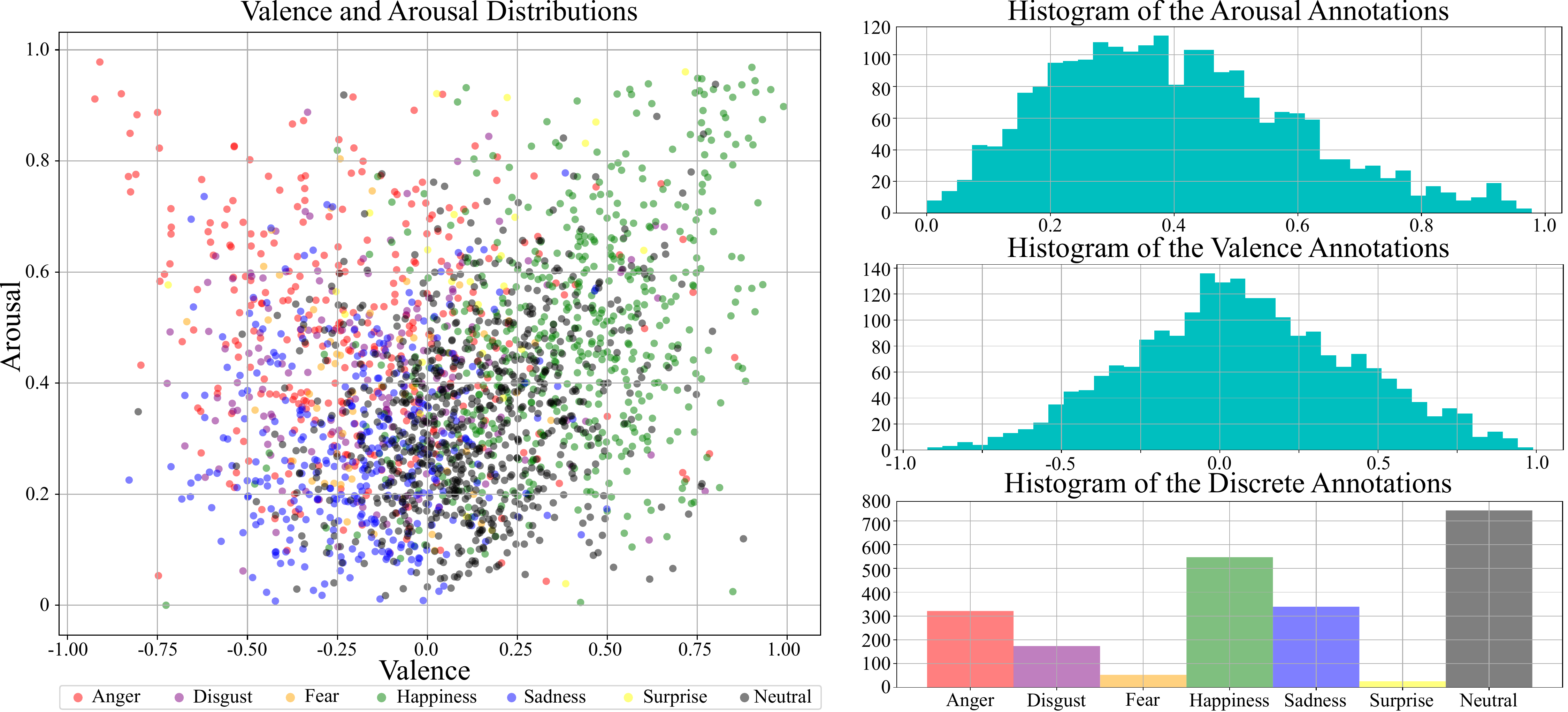}
    \caption{Distribution of utterance-level annotations of the OMG-Emotion Dataset over an arousal/valence scale on the left, and frequencies for arousal, valence and categorical annotations on the right.}
    \label{fig:data_analysis}
\end{figure*}

Our annotation procedure was online for a week, in which we could collect an average of 5.40 annotations per utterance. We then standardized the arousal/valence annotations per utterance. This was important as emotion expressions can be described subjectively using different basis, but the perception variations are maintained. This helps in cases where the annotations were made with subjective starting biases but present a similar transition pattern within utterances. 

Figure \ref{fig:data_analysis} illustrates the annotation distribution. It is possible to see that the categorical annotations were mostly concentrated over neutral expressions, with fearful and surprised expressions being less prevalent. This is common behavior for emotion expressions collected from in-the-wild videos, as we often maintain a neutral demeanor.

It is also possible to see that our arousal/valence annotations are well distributed over the entire spectrum, with more values being annotated as having neutral valence and calm arousal. This is also a common behavior for these types of dataset. Finally, we also provide in Figure \ref{fig:data_analysis} the arousal/valence distribution combined with the categorical annotation, which shows that the arousal/valence annotations are congruent with the categorical labels. 

With these measures, we can show that our corpus is balanced over the arousal/valence space, and presents a challenging task for automatic emotion classification.

\subsection{Gold Standard}
One of the major challenges of emotion recognition lies in attributing an objective metric to the subjective evaluation of emotions. Every individual experiences emotion in a different manner and thus, while annotating the data samples, each annotator introduces their subjectivity when labelling data. This results in different annotations and labels for the same data samples. Therefore, it is difficult to obtain a single label for the data which agrees with all of the annotator's evaluations. One of the ways to deal with this problem is to establish a \textit{gold standard} for the annotations~\cite{Han2017From}. This results in trustworthy labels that are truly representative of the subjective annotations from each of the annotators, providing an objective metric for evaluation. 

The most common approach for achieving this, in the case of categorical or nominal data, is to generate a frequency distribution of all the annotations and take the mode of the histogram to represent the majority vote. In the case of interval or ratio data labels, a mean or median value from all the annotations is computed to represent the gold standard. Although this is an easy and quick solution to the problem, it does not fare well in cases of huge disagreement between authors. Therefore, a more efficient solution involves weighing each of the annotations with respect to their overall agreement with others and computing a weighted estimate of the gold standard for the annotations. 

For the OMG dataset, an Evaluator Weighted Estimator (EWE)~\cite{Grimm2005Evaluation} based approach was used to compute the gold standard evaluation for arousal and valence labels. The resultant EWE label for arousal and valence is computed using equation~\ref{eq:1}.

\begin{equation}
\hat{x}_n^{EWE,(i)} = \frac{1}{\sum_{k=1}^K r_k^{(i)}} \sum_{k=1}^K r_k^{(i)} \hat{x}_{n,k}^{(i)}
\label{eq:1}
\end{equation}

where $\hat{x}_{n,k}^{(i)}$ denotes the annotation provided for the n\textsuperscript{th} sample with $n=1,...,N$ by the k\textsuperscript{th} annotator with $k=1,...,K$ in any of the dimensions $i \in \{Arousal,Valence\}$, and $r_k^{(i)}$ represents the confidence measure assigned to the k\textsuperscript{th} annotator. The confidence for each of the annotators can be computed using equation~\ref{eq:2}.
\begin{equation}
\small r_k^{(i)} = \frac{\sum_{n=1}^N (\hat{x}_{n,k}^{(i)} - \mu_k^{(i)}) (\hat{x}_n^{MLE,(i)}-\mu^{MLE,(i)})}{\sqrt[]{\sum_{n=1}^N (\hat{x}_{n,k}^{(i)} - \mu_k^{(i)})^2} \ \sqrt[]{\sum_{n=1}^N(\hat{x}_n^{MLE,(i)}-\mu^{MLE,(i)})^2}}
\label{eq:2}
\end{equation}
with
\begin{equation}
\small \mu_k^{(i)} = \frac{1}{N} \sum_{n=1}^N \hat{x}_{n,k}^{(i)}
\label{eq:3}
\end{equation}
and 
\begin{equation}
\small \mu^{MLE,(i)} = \frac{1}{N} \sum_{n=1}^N  \hat{x}_{n}^{MLE,(i)}
\label{eq:4}
\end{equation}
where $\hat{x}_n^{MLE,(i)}$ denotes the Maximum Likelihood Estimator for the n\textsuperscript{th} sample given $K$ annotations: 

\begin{equation}
\small \hat{x}_{n}^{MLE,(i)} = \frac{1}{K} \sum_{k=1}^K  \hat{x}_{n,k}^{(i)}
\label{eq:5}
\end{equation}

Thus, for each data sample annotated by at least five annotators, the EWE label is computed for arousal and valence to represent the gold standard. For the emotion labels, represented by seven emotion categories, the maximum vote is used to compute the gold standard.

\subsection{Evaluation Protocol}

To evaluate possible models which could use our dataset in a fair manner, we propose here the use of a standard evaluation protocol for the OMG-Emotion Dataset. First, we separate the data into training and testing subsets. The training set has 118 unique videos and the test set has 60. We then propose two tasks: the first one categorical emotion recognition and the second one continuous estimation, based on arousal and valence values.

The first task of the challenge is to calculate the general F1 score for each utterance:

\begin{equation}
F1 = 2\cdot \frac{precision \cdot recall}{precision + recall}
\label{eq:f1score}
\end{equation}

As we have five labels per category, we calculate the final label by performing a simple max-voting for each utterance. The F1 score gives us a general performance of the model by calculating the harmonic average of the precision and recall, and we consider it more important for this task than a simple accuracy metric.

For the second task, we consider the arousal/valence estimation as a regression problem. Thus, we propose as a metric the computation of the Mean Squared Error (MSE) for arousal and valence values. The MSE gives us a rough indication of how the proposed model is behaving and a simple comparative metric. In addition, we also propose the use of the Concordance Correlation Coefficient (CCC) \cite{Lawrence1989} between the model's output and the annotator's annotations. The CCC can be defined as:

\begin{equation}
ccc = \frac{2 \rho \sigma_x \sigma_y}{\sigma_{x}^2 + \sigma_{y}^2 + (\mu_x - \mu_y)^2}
\label{eq:ccc}
\end{equation}
where $\rho$ is the Pearson's Correlation Coefficient between annotator labels and the gold standard, $\mu_x$ and $\mu_y$ denote the mean for annotator labels and the gold standard and $\sigma_{x}^2$ and $\sigma_{y}^2$ are the corresponding variances.

Both MSE and CCC are calculated based on the model's response and the gold standard. The difference is that the MSE is calculated based on the utterance gold standard, and the CCC based on the video gold standard. This way, while the MSE provides an initial indication of the model's performance, the CCC shows the capability of the model to describe the expressions in a video as a whole, taking into consideration the contextual information.

\section{OMG-Emotion Dataset}

We collected a total of 567 unique videos, totalizing 7371 video clips with each clip consisting of a single utterance. Each video is split into several sequential utterances, each one with an average length of 8 seconds, and having an average length of around 1 minute. After annotation collection, we obtained a total of 39803 unique annotations, averaging 5 annotations per video. Table \ref{tab:omgSummary} exhibits a summary of the OMG-Emotion dataset.

\begin{table}
\caption{Summary information about the OMG-Emotion Dataset.}
    \center \begin{tabular}{| c| c |} \hline
    Videos & 567 (Around 15 hours) \\ \hline
    Utterances & 7371 (12.96 Utterances per video) \\ \hline
    Annotations & 39803 (5.40 Annotations per utterance) \\ \hline    
  \end{tabular} 
\label{tab:omgSummary}
\end{table}

\section{Baseline Experimental Setup}

As a way to foment further development of models based on our dataset, we provide here different benchmarks based on our evaluation protocol. We evaluate the dataset using audio, vision and language-only based models.
As described before, we calculate the utterance-based F1-score for categorical classification, the Mean Squared Error (MSE) for utterance-based arousal/valence estimation and the video-based Congruence Coefficient Correlation (CCC) also for arousal/valence estimation. The arousal/valence estimations are calculated based on the gold standard of all annotators, and the categorical emotion based on a max-voting scheme.


\subsection{Experiments}
\textbf{Audio modality baseline}.
We used the openSMILE IS13-ComParE feature set \cite{schuller2013interspeech}, which includes low-level descriptors like MFCC, voice probability, voice quality, etc, overall comprising a feature vector of 6,373 values. Feature vectors are normalized to have a zero mean and a unit variance. An SVM with an RBF kernel was used as a baseline classifier (SVM A model). 

In addition, we conducted an experiment using the audio channel of a previously proposed work \cite{Barros2016}, pre-trained with the RAVDESS corpus.

\textbf{Text modality baseline}.\par

The baseline text classifier is trained on the OMG training data and uses the word2vec Google News corpus vectors as pretrained word embeddings. The model feeds the transcribed sentences as sequences of word embeddings into a 1D CNN with 128 filters and a kernel size of 3.

The CNN is followed by a global max pooling layer and a fully connected layer with 500 units, ending finally in a single neuron with a sigmoidal activation function that predicts the valence and arousal levels of the input utterance. The CNN and fully connected layer use a ReLU activation and a dropout factor of 0.2 is applied after the word embedding and fully connected layer.

\textbf{Vision modality baseline}.

For our vision experiments, we used a convolution neural network based on the Face-channel proposed on previous works \cite{Barros2016}. To train the network, we use the FER+ dataset.

We perform experiments with the same network architecture, but for two different tasks: one for categorical emotion classification and the other for arousal/valence estimation.

\subsection{Results}
Table \ref{tab:baselineResults} exhibits the summary of our baseline results. It shows the best performance of each of the individual modalities.

\begin{table}
\centering
\caption{Baseline results. }
\begin{tabular}{|l|l|l|l|l|l| }
 \hline 
Model   & Emotion & \multicolumn{2}{|c|}{Arousal} &\multicolumn{2}{|c|}{Valence} \\
      \hline
 & F1-Score & MSE & CCC & MSE & CCC   \\   \hline
 
SVM A &    0.33     &   0.04      &  0.15   &  0.10 &  0.21    \\\hline

RF T  &    0.39     &    0.06     &   0.15      &  0.10   &  0.04 \\\hline

Vision - Face Channel \cite{Barros2016} &   0.37     &   0.05      &   0.12     & 0.12   & 0.23  \\

Vision - Audio Channel \cite{Barros2016} &   0.33     &   0.05      &   0.08     & 0.12   & 0.10  \\
 \hline
 
\end{tabular}
\label{tab:baselineResults}
\end{table}

The OMG-Emotion dataset is a challenging corpus, and this is reflected on the CCC and F1-score values we obtained. Although our experiments are described as baseline, the models and techniques we used are considered state-of-the-art in the field. The MSE by itself does not carry any meaningful information about the performance (only when compared to other models), and thus we report it here for comparison reasons. 

\section{Conclusion and Future Works}
In this paper, we introduce a new emotion recognition dataset with a unique focus on attempting to capture the context of emotion features over a gradual period of time. The considerations taken into account during data collection and annotation remedies many of the issues that occur in other contemporary emotion recognition datasets when attempting to apply these datasets to longer temporal spans. We intend for this dataset to be a unique challenge and a step forward towards more robust and in-depth recognition of emotions in the wild. 

Our dataset will be used on the OMG-Emotion Recognition challenge\footnote{https://www2.informatik.uni-hamburg.de/wtm/OMG-EmotionChallenge/}, where we hope to foster the implementation of neural models for contextual emotion recognition.  We hope to inspire more researchers to pay greater attention to the potential features that can only be reliably tracked over time.

In the future, we hope to increase the number of samples available in the dataset and refine the quality of selected samples. In doing so we hope to increase the robustness and saliency of emotion features provided by the dataset. In particular, we are interested in collecting more samples of currently infrequent emotion samples, such as Fear, Disgust, and Surprise. Furthermore, we hope to refine our baseline by incorporating more advanced neural mechanisms that are well suited to the temporal nature of the dataset.

\section*{Acknowledgements}
The authors gratefully acknowledge partial support from the German Research Foundation DFG under project CML (TRR 169), and the European Union under projects SECURE (No. 642667), and SOCRATES (No. 721619).

\bibliographystyle{IEEEtran}
\bibliography{references/citations.bib}

\begin{thebibliography}{10}
\providecommand{\url}[1]{#1}
\csname url@samestyle\endcsname
\providecommand{\newblock}{\relax}
\providecommand{\bibinfo}[2]{#2}
\providecommand{\BIBentrySTDinterwordspacing}{\spaceskip=0pt\relax}
\providecommand{\BIBentryALTinterwordstretchfactor}{4}
\providecommand{\BIBentryALTinterwordspacing}{\spaceskip=\fontdimen2\font plus
\BIBentryALTinterwordstretchfactor\fontdimen3\font minus
  \fontdimen4\font\relax}
\providecommand{\BIBforeignlanguage}[2]{{%
\expandafter\ifx\csname l@#1\endcsname\relax
\typeout{** WARNING: IEEEtran.bst: No hyphenation pattern has been}%
\typeout{** loaded for the language `#1'. Using the pattern for}%
\typeout{** the default language instead.}%
\else
\language=\csname l@#1\endcsname
\fi
#2}}
\providecommand{\BIBdecl}{\relax}
\BIBdecl

\bibitem{Ekman1971}
P.~Ekman and W.~V. Friesen, ``Constants across cultures in the face and
  emotion.'' \emph{Journal of Personality and Social Psychology}, vol.~17,
  no.~2, p. 124, 1971.

\bibitem{Hamann2012}
S.~Hamann, ``Mapping discrete and dimensional emotions onto the brain:
  controversies and consensus,'' \emph{Trends in cognitive sciences}, vol.~16,
  no.~9, pp. 458--466, 2012.

\bibitem{Izard1992}
C.~E. Izard, ``Basic emotions, relations among emotions, and emotion-cognition
  relations.'' 1992.

\bibitem{Thompson2011}
R.~A. Thompson, ``Methods and measures in developmental emotions research: Some
  assembly required,'' \emph{Journal of Experimental Child Psychology}, vol.
  110, no.~2, pp. 275--285, 2011.

\bibitem{Rani2006}
P.~Rani, C.~Liu, N.~Sarkar, and E.~Vanman, ``An empirical study of machine
  learning techniques for affect recognition in human--robot interaction,''
  \emph{Pattern Analysis and Applications}, vol.~9, no.~1, pp. 58--69, 2006.

\bibitem{Soleymani2017}
M.~Soleymani, D.~Garcia, B.~Jou, B.~Schuller, S.-F. Chang, and M.~Pantic, ``A
  survey of multimodal sentiment analysis,'' \emph{Image and Vision Computing},
  vol.~65, pp. 3--14, 2017.

\bibitem{Ochsner2008}
K.~N. Ochsner and J.~J. Gross, ``Cognitive emotion regulation: Insights from
  social cognitive and affective neuroscience,'' \emph{Current directions in
  psychological science}, vol.~17, no.~2, pp. 153--158, 2008.

\bibitem{Iemocap2008}
C.~Busso, M.~Bulut, C.-C. Lee, A.~Kazemzadeh, E.~Mower, S.~Kim, J.~N. Chang,
  S.~Lee, and S.~S. Narayanan, ``{IEMOCAP}: Interactive emotional dyadic motion
  capture database,'' \emph{Language Resources and Evaluation}, vol.~42, no.~4,
  p. 335, 2008.

\bibitem{mosi2016}
A.~Zadeh, R.~Zellers, E.~Pincus, and L.-P. Morency, ``Mosi: multimodal corpus
  of sentiment intensity and subjectivity analysis in online opinion videos,''
  \emph{arXiv preprint arXiv:1606.06259}, 2016.

\bibitem{emoreact2016}
B.~Nojavanasghari, T.~Baltru{\v{s}}aitis, C.~E. Hughes, and L.-P. Morency,
  ``{EmoReact}: a multimodal approach and dataset for recognizing emotional
  responses in children,'' in \emph{Proceedings of the 18th ACM International
  Conference on Multimodal Interaction}.\hskip 1em plus 0.5em minus 0.4em\relax
  ACM, 2016, pp. 137--144.

\bibitem{Chenetal2017}
O.~O.~R. Weixuan~Chen and R.~W. Picard, ``Gifgif+: Collecting emotional
  animated gifs with clustered multi-task learning,'' in \emph{2017 Seventh
  International Conference on Affective Computing and Intelligent Interaction
  (ACII)}.\hskip 1em plus 0.5em minus 0.4em\relax ACII, 2017.

\bibitem{Zafeiriou2017}
S.~Zafeiriou, D.~Kollias, M.~Nicolaou, A.~Papaioannou, G.~Zhao, and I.~Kotsia,
  ``Aff-wild: Valence and arousal in-the-wild challenge,'' 2017.

\bibitem{Dhall2017}
A.~Dhall, R.~Goecke, S.~Ghosh, J.~Joshi, J.~Hoey, and T.~Gedeon, ``From
  individual to group-level emotion recognition: Emotiw 5.0,'' in
  \emph{Proceedings of the 19th ACM International Conference on Multimodal
  Interaction}.\hskip 1em plus 0.5em minus 0.4em\relax ACM, 2017, pp. 524--528.

\bibitem{eyben2013recent}
F.~Eyben, F.~Weninger, F.~Gross, and B.~Schuller, ``Recent developments in
  opensmile, the munich open-source multimedia feature extractor,'' in
  \emph{Proceedings of the 21st ACM International Conference on
  Multimedia}.\hskip 1em plus 0.5em minus 0.4em\relax ACM, 2013, pp. 835--838.

\bibitem{Han2017From}
J.~Han, Z.~Zhang, M.~Schmitt, M.~Pantic, and B.~Schuller, ``From hard to soft:
  Towards more human-like emotion recognition by modelling the perception
  uncertainty,'' in \emph{Proceedings of the ACM Multimedia Conference}, ser.
  MM '17.\hskip 1em plus 0.5em minus 0.4em\relax New York, NY, USA: ACM, 2017,
  pp. 890--897.

\bibitem{Grimm2005Evaluation}
M.~Grimm and K.~Kroschel, ``Evaluation of natural emotions using self
  assessment manikins,'' in \emph{IEEE Workshop on Automatic Speech Recognition
  and Understanding, 2005.}, Nov 2005, pp. 381--385.

\bibitem{Lawrence1989}
L.~I. Lin, ``{A} concordance correlation coefficient to evaluate
  reproducibility,'' \emph{Biometrics}, vol.~45, no.~1, pp. 255--268, Mar 1989.

\bibitem{schuller2013interspeech}
B.~Schuller, S.~Steidl, A.~Batliner, A.~Vinciarelli, K.~Scherer, F.~Ringeval,
  M.~Chetouani, F.~Weninger, F.~Eyben, E.~Marchi \emph{et~al.}, ``The
  interspeech 2013 computational paralinguistics challenge: social signals,
  conflict, emotion, autism,'' in \emph{Proceedings INTERSPEECH 2013, 14th
  Annual Conference of the International Speech Communication Association,
  Lyon, France}, 2013.

\bibitem{Barros2016}
P.~Barros and S.~Wermter, ``Developing crossmodal expression recognition based
  on a deep neural model,'' \emph{Adaptive Behavior}, vol.~24, no.~5, pp.
  373--396, 2016.

\end{thebibliography}

\end{document}